\font\boldsym=cmmib10\font\boldsym=cmmib10

\def\Omegab{{\hbox{\boldsym\char'012}}}

\def\nub{{\hbox{\boldsym\char'027}}}
\def\csib{{\hbox{\boldsym\char'030}}}
\def\pib{{\hbox{\boldsym\char'031}}}

\def\omegab{{\hbox{\boldsym\char'041}}}

\catcode`\@=11
\def\gsim{\ifmmode{\mathrel{\mathpalette\@versim>}}
    \else{$\mathrel{\mathpalette\@versim>}$}\fi}
\def\lsim{\ifmmode{\mathrel{\mathpalette\@versim<}}
    \else{$\mathrel{\mathpalette\@versim<}$}\fi}
\def\@versim#1#2{\lower 2.9truept \vbox{\baselineskip 0pt \lineskip
    0.5truept \ialign{$\m@th#1\hfil##\hfil$\crcr#2\crcr\sim\crcr}}}
\catcode`\@=12
\def\spose#1{\hbox to 0pt{#1\hss}}
\def\lta{\mathrel{\spose{\lower 3pt\hbox{$\mathchar"218$}}
     \raise 2.0pt\hbox{$\mathchar"13C$}}}
\def\gta{\mathrel{\spose{\lower 3pt\hbox{$\mathchar"218$}}
     \raise 2.0pt\hbox{$\mathchar"13E$}}}

\def\msun{{\,M_\odot}}
\def\lsun{L_{\odot}}

\def\Av{{\bf A}}

\def\alst{\alpha_*}
\def\alsn{\alpha_{\rm SN}}

\def\csiv{\csib}
\def\csici{\xi_i}

\def\csicu{\xi_1}
\def\csicd{\xi_2}
\def\csict{\xi_3}

\def\ddn{d^2{\bf n}}
\def\dtv{d^3{\bf v}}
\def\dtx{d^3{\bf x}}
\def\deg{^\circ}
\def\Deltatp{\Delta t_{\rm p}}

\def\eps{\epsilon}
\def\evi{{\bf e}_i}

\def\Ez{{\cal E}_0}

\def\gv{{\bf g}}
\def\gvg{\gv_{\rm g}}
\def\gvp{\gv_{\rm p}}
\def\gvc{\gv_{\rm cl}}
\def\gradx{\nabla_{\bf x}}
\def\gradcsi{\nabla_{\csiv}}
\def\gradR{\nabla_{\bf R}}

\def\Lb{L_{\rm B}}
\def\Lx{L_{\rm X}}
\def\Lsn{L_{\rm SN}}
\def\Lstar{L_{\star}}

\def\mas{M _*}
\def\mah{M _{\rm h}}
\def\map{M_{\rm p}}
\def\mag{M_{\rm g}}
\def\mcl{M_{\rm cl}}

\def\nv{{\bf n}}
\def\nci{n_i}
\def\nuv{\nub}
\def\nuci{\nu_i}

\def\Om{{\cal O}}

\def\Omt{\Om^{\rm T}}
\def\Ome{\Omegab}
\def\dOme{\dot\Ome}
\def\ome{\omegab}

\def\pv{{\bf p}}
\def\pgv{\pib}
\def\pci{p_i}

\def\phic{\phi_{\rm cl}}
\def\phig{\phi_{\rm g}}
\def\phip{\phi_{\rm p}}
\def\phis{\phi_*}
\def\phih{\phi_{\rm h}}

\def\ros{\rho _*}
\def\roh{\rho _{\rm h}}
\def\rog{\rho_{\rm g}}
\def\rop{\rho_{\rm p}}
\def\rocl{\rho_{\rm cl}}
\def\roclm{\overline{\rocl}}
\def\rosz{\rho _{*0}}
\def\rohz{\rho _{\rm h0}}
\def\rcs{r_{\rm *}}
\def\rch{r_{\rm h}}
\def\rcl{r_{\rm cl}}
\def\rt{r_{\rm t}}
\def\Rv{{\bf R}}
\def\Rci{R_i}
\def\Rcj{R_j}
\def\Rp{\Rv_{\rm p}}
\def\rpv{{\bf r}_{\rm p}}
\def\rp{r_{\rm p}}
\def\rpmin{\rp^{\rm min}}
\def\Rsn{R_{\rm SN}}

\def\sourceM{{\cal M}}
\def\sourcePi{{\cal P}_i}
\def\sourcePv{{\bf P}}
\def\sourceE{{\cal E}}
\def\sourceK{{\cal K}}
\def\sourceL{{\cal L}}
\def\sigij{\sigma^2_{ij}}
\def\SX{\Sigma_{\rm X}}
\def\sigcl{\sigma_{\rm cl}}

\def\tcc{t_{\rm cc}}
\def\Tr{{\rm Tr}}
\def\Tcl{T_{\rm cl}}
\def\Tp{T_{\rm p}}
\def\tm{t_{\rm m}}

\def\tip{t^{\rm i}_{\rm p}}
\def\tH{t_{\rm H}}

\def\us{u_{\rm s}}
\def\uv{{\bf u}}
\def\uci{u_i}
\def\uw{u_*}
\def\usn{u_{\rm SN}}

\def\vv{{\bf v}}
\def\vci{v_i}
\def\vcj{v_j}
\def\vm{\overline{\vv}}
\def\vmci{\overline{v}_i}
\def\vmcj{\overline{v}_j}
\def\Vv{{\bf V}}

\def\xv{{\bf x}}
\def\xci{x_i}

\documentstyle[11pt,aaspp4]{article}

\slugcomment{In press on ApJ (Main Journal)}

\lefthead{D'Ercole, A., Recchi, S., Ciotti, L.}
\righthead{Tidal interactions on Es gas flows}

\begin{document}

\title{Effects of Tidal Interactions\\
       on the Gas Flows of Elliptical Galaxies}

\author{A. D'Ercole$^{1}$, S. Recchi$^{2,3}$}
\and
\author{L. Ciotti$^{1,4}$}

\affil{$^{1}$Osservatorio Astronomico di Bologna, via Ranzani 1, 
             40127 Bologna, ITALY}
\affil{$^{2}$Dip. di Astronomia, Universit\`a di Bologna, 
             via Ranzani 1, 40127 Bologna, ITALY}
\affil{$^{3}$SISSA/ISAS, via Beirut 2-4, 34014 Trieste, ITALY}
\affil{$^{4}$Scuola Normale Superiore, Piazza dei Cavalieri 7,
            56126 Pisa, ITALY}



\begin{abstract}
During a Hubble time, cluster galaxies may undergo several mutual
encounters close enough to gravitationally perturb their hot, X--ray
emitting gas flows.  We ran several 2D, time dependent hydrodynamical
models to investigate the effects of such perturbations on the gas
flow inside elliptical galaxies, focusing on the expected X--ray
properties.  In particular, we studied in detail the modifications
occurring in the scenario proposed by D'Ercole et al. (1989), in which
the galactic interstellar medium produced by the aging galactic
stellar population, is heated by type Ia supernovae (SNIa) at a
decreasing rate.  We find that, although the tidal interaction in our
models lasts less than 1 Gyr, its effect extends over several Gyrs.
The tidally induced turbulent flows create dense filaments which cool
quickly and accrete onto the galactic center, producing large spikes
in the global X--ray luminosity $\Lx$. Once this mechanism starts, it
is fed by gravity and amplified by SNIa.  This evolution is found to
be virtually independent of the dynamical state of the gas flow at the
beginning of the interaction.  To better understand the role of SNIa
heating, we also considered a ``pure'' cooling flow model without
supernovae; in this case the amplitude of the $\Lx$ fluctuations due
to the tidal interaction is substantially reduced.  We conclude that,
if SNIa significantly contribute to the energetics of the gas flows in
ellipticals, then the observed spread in the $\Lx -\Lb$ diagram at any
fixed optical galaxy luminosity $\Lb\gta 3\times 10^{10}\lsun$ may be
caused, at least in part, by this mechanism.  On the contrary, tidal
interactions cannot be responsible for the observed spread if the pure
cooling flow scenario applies.
\end{abstract}



\section{INTRODUCTION}
In a previous paper D'Ercole et al. (1989) proposed the
wind--outflow--inflow (WOI) scenario as a possible explanation for one
of the most striking properties of the X--ray emission of elliptical
galaxies, i.e., the large scatter in the $\Lx -\Lb$ diagram of roughly
two orders of magnitude in $\Lx$ at any fixed $\Lb\gta 3\times
10^{10}\lsun$ (Fabbiano 1989, Fabbiano, Kim \& Trinchieri 1992).  In
their subsequent extensive exploration of 1D hydrodynamical models
Ciotti et al. (1991, hereafter CDPR) assumed that the type I supernova
(SNIa) rate decreases in time as $\Rsn\propto t^{-1.5}$, faster than
the decrease of the mass return rate from the (passively) evolving
stellar population, which varies in time approximately as $t^{-1.36}$.
In these models, the gas lost by the stars is initially driven out of
the galaxy through a supersonic wind powered by the thermalization of
the kinetic energy injected in the interstellar medium (ISM) by the
SNIas.  As the specific energy input decreases, the wind turns into a
subsonic outflow and eventually the flow reverts to an inflow regime,
at the time of the so--called {\it cooling catastrophe} ($\tcc$).  In
pace with the content of the X--ray emitting hot gas, $\Lx$ decreases
during the wind phase and increases during the outflow regime up to a
maximum reached at $\tcc$.  Later on, $\Lx$ slowly decreases roughly
following the quenching of the SNIa heating $\Lsn\propto\Rsn$.  CDPR
ran several 1D numerical simulations, showing how $\tcc$ critically
depends on the values of the structural parameters of the host galaxy,
such as the scale--length of the stellar distribution (e.g, the
effective radius for de Vaucouleurs profiles or the so--called core
radius for King profiles), the scale--length of the dark matter halo,
the relative amount of total masses of the two distributions.  Small
changes in one (or more) of these parameters produce a large shift (of
the order of some Gyr) of $\tcc$, causing two galaxies of virtually
the same $\Lb$ to be caught at very different $\Lx$ after a Hubble
time. As a consequence, the above--mentioned $\Lx$ scatter at fixed
$\Lb$ finds a simple explanation in the WOI scenario.

The main objection raised against this scenario has been the claimed
detection of low iron abundance in the ISM of ellipticals.  In fact,
under the assumption of solar abundance ratios, the analysis of the
available data suggests a very low iron abundance, consistent with no
SNIas enrichment, and even lower than that of the stellar component
(Ohashi et al. 1990, Awaki et al. 1991, Ikebe et al. 1992, Serlemitsos
et al.  1993, Loewenstein et al. 1994, Awaki et al. 1994, Arimoto et
al. 1997, Matsumoto et al. 1997). However, some authors have found
that more complex multitemperature models with a higher abundance give
a better fit to the data (Kim et al. 1996, Buote \& Fabian 1998).
This is rather puzzling, in that a value as high as one fourth of the
standard Tammann (1982) rate $\Rsn=0.22$ SNU (for $H_0=50$ km s$^{-1}$
Mpc$^{-1}$) agrees with the current optical estimates of the present
day SNIa rate (Cappellaro et al. 1997). Several models have been
calculated with this new rate (Pellegrini \& Ciotti 1998, D'Ercole \&
Ciotti 1998, hereafter DC), resulting in the main features of the WOI
scenario being preserved, provided the amount of dark matter is
properly scaled down.

In any case, even though $\Rsn\approx 0$, it is well known that the
pure cooling flow scenario is unable, in isolated galaxies, to
reproduce the observed scatter (see, e.g., CDPR). As a consequence,
the possible environmental origin of the $\Lx -\Lb$ scatter has been
proposed.  Although some discussion was dedicated to environmental
effects, the bulk of the CDPR simulations did not consider ambient
influences. It is known, however, that the majority of early type
galaxies is found in clusters, where they certainly suffer several
environmental interactions. More specifically, 1) thermal conduction
with the intracluster medium (ICM); 2) ram pressure stripping of the
ISM due to the motion of the galaxies with respect to the ICM; 3)
influence of the cluster tidal field; 4) galaxy--galaxy gravitational
encounters.

Because the ICM in rich clusters is much hotter ($T\simeq 10^8$ K)
than the galactic gas flows ($T\simeq 10^7$ K), heat is expected to
flow into the galactic ISM if thermal conduction is not suppressed by
magnetic fields.  It is easy to show that the heat transferred by an
unimpeded thermal conduction is much larger than any other galactic
energy budget, and evaporation can hardly be avoided. The very
existence of bright X--ray ellipticals thus suggests that heat
conduction must be suppressed to a great extent (see, e.g., CDPR, and
references therein).

The ram pressure effect of the ICM on the ISM of the elliptical
galaxies was discussed by D'Ercole et al. (1993) in the WOI scenario,
and, e.g., by Balsara, Livio \& O'Dea (1994, and references therein)
in the cooling flow scenario.  While numerical models show that ram
pressure is very effective in stripping galaxies of gas, observations
reveal that numerous cluster galaxies are strong X--ray emitters, thus
showing their ability to retain their hot ISM. For example, Dow \&
White (1995), found that ellipticals in Coma are as X--ray bright as
in Virgo.  This situation appears rather puzzling, and certainly not
satisfactorily understood.

The mean cluster gravitational field acts on the structure of each
galaxy and its gaseous halo, as a weak tidal field.  $N$--body
numerical simulations (e.g., Ciotti \& Dutta, 1994) show that its
effect on the galactic stellar component is very small, consisting
essentially in a re--orienting of the galactic major axis with respect
to the cluster center.

In summary, while some effort was spent by several authors in
considering the influence of the ICM on the ISM of ellipticals, up to
now no attempt was made to study the response of the X--ray emitting
ISM of cluster ellipticals to the tidal interaction due to the cluster
field and to a close encounter with a nearby galaxy. As we will see,
in a typical cluster, encounters close enough to have non negligible
dynamical effects on the ISM are likely to occur.

In this paper we show the results of 2D numerical hydrodynamical models
focusing on the influence both of the cluster tidal field and of a
transient external gravitational field on the evolution of the ISM of
ellipticals in the WOI and cooling flow scenarios.  In \S 2 we
describe the galaxy model, the hydrodynamical equations, and the
assumptions adopted in our simulations. The results are presented in
\S 3. A discussion and the conclusions are finally given in \S 4. 

\section{The Galaxy Model}
In order to better compare the present results with those obtained by
CDPR, we adopt for the model galaxy the same bright and dark matter
(DM) distributions considered by these authors for their basic model
(the so--called {\it King Reference Model}, KRM).  The stellar density
distribution is a truncated King (1972) model
\begin{equation}
\ros(r)={\rosz\over (1+r^2/\rcs^2)^{3/2}},
\quad r\leq\rt ,
\end{equation}
and the DM density profile is a truncated quasi--isothermal model
\begin{equation}
\roh (r)={\rohz\over 1+r^2/\rch^2},
\quad r\leq\rt .
\end{equation}
Here $(\rosz,\rcs)$ and $(\rohz,\rch)$ are the central density and
scale--length of the luminous and DM density distributions,
respectively, while $\rt$ is their common truncation radius.  The total
galaxy density, potential, and mass are $\rog=\ros+\roh$,
$\phig=\phis+\phih$, and $\mag=\mas+\mah$.

As for the KRM in CDPR we assume $\rosz =6.08\times 10^{-21}$ g
cm$^{-3}$, $\rohz =7.6\times 10^{-23}$ g cm$^{-3}$, $\rcs =368$ pc,
$\rch=4.5\rcs$ and $\rt=66.24$ kpc.  With these parameters
$\mas=2.75\times 10^{11}\msun$ and $\mah=2.48\times 10^{12}\msun$,
respectively, and the galaxy lies on the so--called Fundamental Plane.
It should be noted that several recent indications, from
high--resolution $N$--body simulations (Dubinski \& Carlberg 1991;
Navarro, Frenk \& White 1997) as well as from theoretical arguments
(Evans \& Collett 1997; Ciotti 1996,1999), seem to suggest a peaked
density distribution for the DM halos, as for example that used in the
numerical simulations of gas flows carried out by Pellegrini \& Ciotti
(1998).  In spite of this, we decided to maintain here the functional
form for the halo as in the KRM, in order to better understand the
effects of the tidal interactions by comparison with the unperturbed
model.

Of course, a self--consistent treatment of the evolution of galactic
gas flows in the cluster environment is a formidable task, both for
the complexity of the involved physics, and for the high
dimensionality of the parameter space (requiring very large
computational resources to be fully explored).  Our aims are far more
narrow, focusing on the qualitative understanding of the effects of a
representative tidal encounter. In this line, the mass $\map$ of the
perturbing galaxy (the {\it perturber}) is assumed to be equal to the
mass of the galaxy, $\map=\mag\simeq 2.8\times 10^{12}\msun$, because
we are not interested here in the effects of interactions with
significantly less massive galaxies.  In a subsequent paper
(Pellegrini, D'Ercole \& Ciotti, in preparation) we explore the
effects produced on the X--ray emission of a galaxy model similar to
that here discussed by a collision with a significantly less massive
cluster galaxy.

\subsection{Assumptions and Equations}
We derive and discuss here the hydrodynamical equations describing the
evolution of the gas flows in our models, using the general equations
given in Appendix A.2. In the following, bold--face symbols represent
vectors.  In the (baricentric) cluster inertial reference frame $S$,
the center of mass of the galaxy and of the perturber are $\Rv$ and
$\Rp$, respectively; moreover, $\rpv=\Rp-\Rv$ and $\rp=||\rpv||$,
where $||\cdot ||$ is the standard norm.  For computational reasons
these equations are written in a non--inertial reference system $S'$
fixed by the following rules: the origin of $S'$ coincides with $\Rv$,
$\csiv$ is the position vector in $S'$, the $\csict$ axis points at
all times towards $\Rp$, and finally $r=||\csiv||$.

\subsubsection{The Effective Gravitational Field}
We derive first the effective gravitational field experienced by the
gas in the frame $S'$ (see equation [A16]), and we discuss two
assumptions used in our models, namely that 1) the cluster tidal field
is negligible during the encounter, and that 2) the perturber
trajectory can be approximated by a straight line, covered with
constant velocity. Although in the numerical code the gravitational
forces are not approximated by their linear expansion (tidal regime),
here we consider the associated tidal fields, in order to obtain the
geometrical conditions for the validity of the assumptions above.

At any position $\xv$ in the cluster frame $S$, the total acceleration
is $\gv=-\gradx(\phic+\phip+\phig)$, where $\phic$, $\phip$, and
$\phig$ are the cluster, perturber, and galaxy potentials,
respectively, and $\gradx$ is the gradient operator with respect to
$\xv$.  Note that the self--gravity of the gaseous component is not
considered.  The acceleration $\Av$ of the center of mass of the
galaxy is given by $\mag\Av=-\int\rog\gradx(\phic+\phip)\dtx$.  In the
tidal approximation $\phic$ and $\phip$ are expanded up to the second
order around $\Rv$, i.e.,
\begin{equation}
\phic(\xv)\sim \phic(\Rv)-
                <\gvc(\Rv),\xv-\Rv>-
                {<\Tcl(\Rv)(\xv-\Rv),\xv-\Rv>\over 2},
\end{equation}
where $\gvc(\Rv)=-\gradR\phic(\Rv)$, $<,>$ is the standard inner
product, and $\Tcl$ is the tidal tensor associated with $\phic$, whose
elements are given by
\begin{equation}
T_{\rm cl,ij}(\Rv)=-{\partial^2\phic(\Rv)\over\partial\Rci\partial\Rcj},
                   \quad (i,j=1,2,3).
\end{equation}
A similar relation holds for the gravitational field produced by the
perturber, where now $\gvp(\Rv)=-\gradR\phip(\Rv)$ and $\Tp$ is the
tidal tensor associated with $\phip$.  In this approximation, simple
calculations show that the effective gravitational field entering
equation (A16) is given by
\begin{equation}
\gv-\Av\sim [\Tcl(\Rv)+\Tp(\Rv)](\xv-\Rv)+\gvg,
\end{equation}
where $\gvg=-\gradx\phig$. 
The coordinate transformation from $S$ to $S'$ shows that
\begin{equation}
\Omt\gvg=-G{\mag(r)\over r^3}\csiv,
\end{equation}
and 
\begin{equation}
\Omt [\Tp(\Rv)(\xv-\Rv)]=-{G\map\over\rp^3}(\csicu ,\csicd , -2\csict ).
\end{equation}
Equations (6) and (7) are derived under the assumption of spherical
mass distributions. Actually, the galaxy and the perturber will change
their shape during the interaction.  However, for small deformations
the potential remains rounder than the underlying density distribution
(see, e.g., Binney \& Tremaine 1987, hereafter BT), and so we regard
equation (6) as still largely valid for our problem.  Moreover, the
effect of the deformation of the pertuber density influences the tidal
tensor $\Tp$ given in equation (7) only at higher order in $\rp$ (see,
e.g., Danby 1962, Chapter 5).  As a consequence, in all simulations
the (visible and dark) mass densities of the galaxy and of the
pertuber are assumed to maintain their initial spherical symmetry.

As anticipated in point 1) above, in our simulations we do not
consider the term $\Tcl$ appearing in equation (5), in order to reduce
the dimensionality of the parameter space.  The influence of the
cluster tidal field in absence of close encounters was explored
separately, and it is discussed in \S 4.  We now determine when the
perturber tidal field $\Tp$ is dominant over $\Tcl$.  The cluster
tidal field at distance $R=||\Rv||$ from the cluster center is given
by equation (4), and its explicit expression is
\begin{equation}
\Tcl(\Rv)=-{4\pi G\roclm (R)\over 3}\left(\matrix{ 1 & 0 & 0\cr
                                                   0 & 1 & 0\cr
                                                   0 & 0 & 3q -2\cr}
                                    \right),
\end{equation} 
where $\roclm (R)=3\mcl (R)/4\pi R^3$ is the cluster mean density inside $R$, 
and $0\leq
q(R)=\rocl (R)/\roclm (R)\leq 1$. The matrix expression of $\Tcl$ is
obtained in the reference frame $S$ with its $z$ axis aligned along $\Rv$ (see, e.g., Ciotti
\& Dutta 1994). 
As a consequence, in equation (8) the {\it radial} and the {\it
tangential} components of the cluster tidal field are
apparent\footnote{Note that the tidal field outside a spherical mass
distribution given in equation (7) can be obtained directly from
equation (8) for $q=0$, and by substitution of $\rocl$ with $\rop$.}.
In order to obtain a simple estimate of the relative importance of the
cluster and pertuber tidal fields, we consider a typical cluster with
a density profile $\rocl$ as that given in equation (1), with a core
radius $\rcl=350$ kpc, and a central velocity dispersion $\sigcl=10^3$
km s$^{-1}$ (Sarazin 1986). We compare the {\it minimum} (absolute)
value of the perturber tidal field components, $G\map/\rp^3$, with the
radial and tangential components of the cluster tidal field.  In
Fig. 1 these quantities are showed, and it turns out that the cluster
tidal field can be neglected for $\rp\simeq 100$ kpc everywhere but
inside the cluster core.

\placefigure{fig1}

\subsubsection{The Pertuber Trajectory}
Following the arguments above, we assumed in our simulations
$\rpmin=100$ kpc as the minimum distance during the encounter, in
order to safely neglect the cluster tidal force on the ISM.  Actually,
the (full) perturber gravitational field is taken into account for
distances much larger than $\rpmin$, i.e., for $\rp\leq 450$ kpc.
This in order to avoid a sudden change in the total gravitational
field experienced by the gas flow, that could be at the origin of
unphysical oscillations.  In other words, up to the time $\tip$ (a
free parameter in the simulations, whose choice is given in \S 3) the
gas flow evolves as in CDPR, successively the perturber is placed at a
distance of $450$ kpc, with an initial velocity of $V=1000$ km
s$^{-1}$ relative to the galaxy.  We indicate with $\tm$ the time at
which $\rp(\tm)=\rpmin$, and with $\Deltatp=\tm-\tip$.

In order to follow the time evolution of the ISM, we obviously need to
know how $\rp$ changes with time. Adopting equation (3) to
describe the effect of the cluster gravitational field on the galaxy
and pertuber motions, one obtains
\begin{equation}
\ddot\rpv\sim -G(\mag+\map){\rpv\over\rp^3}+\Tcl(\Rv)\rpv.
\end{equation}
With the assumed parameters for the galaxy, the perturber, and the
cluster density distribution, it is easy to verify from the equation
above that the effects of $\Tcl$ on the motion of the two galaxies are
non negligible for values of $\rp$ in the range of interest, and so
the relative orbit cannot be described by a two--body problem.  As
anticipated by point 2) in \S 2.1.1, we arbitrarily adopt a linear
trajectory for $\rpv$, described at constant velocity $V$.  In this
assumption $\Deltatp\simeq 460$ Myr, and simple geometrical arguments
show that the only non--zero component of the angular velocity in
$S'$, entering equation (A16), is given by:
\begin{equation}
|\Omega_1|={\rpmin V\over V^2(t-\tm)^2+(\rpmin)^2}.
\end{equation}
We point out that, as a consequence of the geometry of the problem,
the non--inertial forces associated with the rotation of $S'$ cannot be
taken into consideration by our 2D code, and so we drop these terms in
the hydrodynamical equations used in the numerical simulations
(equations [12]-[14]). In \S 4 we will calculate {\it a posteriori},
from the computed models, the ratio between the gravity and the
fictitious forces associated with $\Ome$, in order to obtain a rough
estimate of the severity of this approximation.  Although this
assumption may appear rather rough, the computed models show that the
perturbations in the ISM do not depend on the details of the physical
process which generate them.  Thus our assumption is not particularly
severe.

\subsubsection{The Source Terms}
As already discussed in the literature (Mathews \& Baker 1971), the
mass and energy sources in elliptical galaxies are associated with their
evolving stellar population: the two main contributions are due to the
gas lost by the red giants and to the ejecta of SNIas. The total mass
return rate, expressed in the general case in equations (A15)-(A17),
is given here by $\sourceM=\alpha(t)\ros$, where $\alpha(t)=
\alst(t)+\alsn(t)$. We adopt the same $\alpha(t)$ as 
in CDPR (equations [3] and [7] there). In particular, this corresponds
to a present day $\Rsn=0.15$ SNU, i.e., 2/3 of the standard Tammann's rate. 
Moreover, we assume that the internal energy 
associated with the mass ejection in both processes is negligible with 
respect to the thermalization
of the stellar velocity dispersion and to the SNIa ejection velocity.
The {\it exact} expression for the {\it injection energy} is given by
\begin{equation}
\Ez={\alst\uw^2+\alsn\usn^2\over 2\alpha} +{\Tr (\sigma^2)\over 2}
     \simeq {\alsn\usn^2 +\alst\Tr (\sigma^2)\over 2\alpha},
\end{equation}
where the last formula is derived under the hypothesis that 1) the
velocity of stellar winds is considerably lower than the stellar
velocity dispersion (i.e., $\uw^2<<\sigma^2$), and that 2) the stellar
velocity dispersion is lower than the velocity of SNIa ejecta (i.e.,
$\sigma^2 <<\usn^2$).  In accordance with the previous works we adopt
the second, approximate expression for $\Ez$, but we remind that in
some circumstances the full one should be used.  The cooling rate per
unit volume is given by $\sourceL=n_{\rm e}n_{\rm p}\Lambda(T)$, where
we parametrize $\Lambda(T)$ following the prescription of Mathews \&
Bregman (1978), after correction of a wrong sign in their equation
(A2).  Finally, no streaming velocity of the stellar component is
assumed to be present in $S'$, and accordingly in the momentum and
energy equations (A16)-(A17) we set $\vm '=0$. This assumption can be
qualitatively explained considering that, when the mutual interaction
between the galaxy and the perturber starts, their stellar and dark
matter distributions are distorted. These deformations are responsible
for the appearance of a net torque acting on the galaxy density
distribution, from which an angular momentum is originated. Of course,
the real situation is much more complex, but our simple assumptions
should capture the kinematical behavior of the galaxy stellar and dark
matter components.

\subsubsection{The Hydrodynamical Equations}
As a consequence of all the previous assumptions and discussions,
equations (A15)-(A17) are specialized in the numerical code as
follows:

\begin{equation}
{\partial\rho\over\partial t}+\gradcsi\cdot\rho\nuv=\alpha\ros,
\end{equation}
\begin{equation}
{\partial\rho\nuv\over\partial t}+\gradcsi\cdot(\rho\nuv\otimes\nuv)
=\rho(\gvg+\Tp\csiv)-(\gamma-1)\gradcsi E,
\end{equation}
\begin{equation}
{\partial E\over\partial t}+\gradcsi\cdot E\nuv=-
                             (\gamma-1)E\gradcsi\cdot\nuv+
                             {\alpha\ros\over 2}||\nuv||^2+
                             \alpha\ros\Ez-
                             \sourceL.
\end{equation}
$\rho$, $E$, and $\nuv$ are respectively the density, the internal
energy per unit volume, and the velocity of the gas, expressed in
$S'$. The explicit expressions of $\gvg$ and $\Tp\csiv$ in this frame
are given in equations (6) and (7).  The gas pressure is
$p=(\gamma-1)E$, where $\gamma=5/3$ is the ratio of the specific
heats. To integrate the set of equations we used a second--order,
upwind, 2D numerical code described in DC, coupled with a staggered,
spherical, Eulerian grid.  Given the symmetry of the problem, the grid
covers polar angles $0\deg\leq\theta\leq 180\deg$ with 82 equally
spaced angular zones. The radial coordinate is divided into 72 zones
and extends up to 95 kpc. In order to get a reasonable spatial
resolution in the inner region, a geometrical progression is adopted
for the radial mesh size, the first one being 100 pc wide. Reflecting
boundary are assumed everywhere but at the outer boundary, where
outflow conditions are adopted. With the adopted non--inertial
reference frame $S'$ and numerical grid, the perturber is always
outside the grid and moves along the $\theta=0$ direction (i.e., along
the $\csict$ axis).  As in CDPR, the gas temperature is not allowed to
drop below $10^4$ K, to avoid an excessive reduction of the time step
when a rapid cooling is present.  We assume the model galaxy to be
initially devoid of gas due to the previous activity of the Type II
SNe. A discussion on the reliability and implications of this
assumption is given in CDPR.

\section{The Models}
Given the Courant condition on the angular mesh close to the center,
the time steps for the numerical computations are rather short (few
10$^3$ yr). We thus computed only few models.  We first ran a model
without the perturber and cluster gravitational field (hereafter model
M0), i.e., a model analogous to the KRM presented in CDPR. As for the
KRM, the cooling catastrophe occurs before 15 Gyr, and all the three
dynamical phases -- wind, outflow and inflow -- are
recovered\footnote{Note however that in model M0 $\tcc$ is greater
than in the KRM (11 Gyr instead of 9.3 Gyr, see Fig. 3). This depends
on the different value assumed for the mean molecular weight (0.62
instead of 0.5) which enters quadratically in the cooling term. By the
way, this shows once more the extreme dependence of the WOI models on
the parameter values.}.

In order to understand whether there are differences in the flow
evolution depending on the time $\tm$ at which the maximum approach
happens, we ran three models (M1, M2, M3) for three different values
of $\tm$, namely $\tm^{1,2,3}=(6.3,10.8,12.6)$ Gyr.  In each of these
models the initial conditions are given by the hydrodynamical
quantities of model M0 at $\tip=\tm-\Deltatp$. In model M1 the galaxy
starts to be perturbed when it is still in its wind phase, while in
models M2 and M3 the galaxy is perturbed when is in the outflow and
inflow phase, respectively (see Fig. 3).  We also ran a pure cooling
flow model (hereafter model MC) in which the energy injection by SNIa
is absent. For this model we set $\tip=4.9$ Gyr.  Although this model
is expected to be rather similar to model M3, we ran it to ascertain
whether the presence of the SNe energy input may affect the perturbed
gas flow, even in the inflow phase.

Finally, we ran two more models without perturber, in order to check
the sensitivity of our results to some of the assumptions made.
Namely, 1) model MP is analogous to model M3, but with a {\it random}
large--scale velocity field perturbation added to the initial
conditions, as done by Kritsuk, B\"ohringer \& M\"uller (1998) in
their simulations; and 2) model MT is analogous to model MC, but with
the cluster tidal field taken into account during a galaxy orbit
inside the cluster.

\subsection{Models Discussion}
We discuss here in detail the M1 model, in which the galaxy starts to
be perturbed during the wind phase. The other M2 and M3 models are
presented by comparison with the M1 model, pointing out differences and
analogies.

\placefigure{fig2}

\subsubsection{Hydrodynamics}

Figure 2 shows the density contours and the velocity field of the gas
at four different times. Panel $a$ is a shot taken at a time very
close to $\tm$, when the distance to the perturber is slightly larger
than $\rpmin$; the exact time and relative distance are $(t,\rp)$=(6.4
Gyr, 141.4 kpc).  The ISM in the galactic region facing the perturber
accelerates toward the right direction (where the perturber is
located), and is lost by the galaxy at higher velocity than in the
opposite direction.  This apparent asymmetry is due to the fact that
in the code the full gravitational field of the perturber is
considered, which, for short distances, differs significantly from the
associated (symmetric) tidal field. In panel $b$ ($t=9.9$ Gyr) the
effects of the encounter are still clearly visible, although the
perturber distance from the galaxy (for the assumed rectilinear
trajectory) is $\rp=3.6$ Mpc.  This is also true at later times (panels
$c$ and $d$), where eddies are established and different dynamical
regimes are present, some regions being in outflow and others in
inflow, a situation analogous of that found in DC on their
investigation on gas flows in (isolated) S0 galaxies.

\subsubsection{X--ray Luminosity and Gas Mass Evolution}
The X--ray luminosity evolution of model M1 is shown as a solid line
in Fig. 3a. For comparison, the dashed line represents the $\Lx$ temporal
evolution of the model M0, and the dotted line shows the time
evolution of the SNIa energy input for unit time, $\Lsn$.

\placefigure{fig3}

As the encounter with the perturber occurs, the cooling catastrophe is
anticipated, as a consequence of the increase in the mean density in
the galactic central regions produced by the tidal perturbation, and
$\Lx$ reaches its (first) maximum at $\tcc\simeq 9.4$ Gyr, earlier
than for model M0 ($\tcc\simeq 11$ Gyr).  This is an interesting
result.  In fact, note that considering spherical coordinates with the
$\csict$ polar axis ($\theta=0\deg$) oriented towards the perturber
(the coordinate system actually adopted in the numerical code), from
equation (7) one obtains that the radial component of the tidal force
(for unit mass) is given by $F_{\rm rad}=-(G\map
r/\rp^3)(\sin^2\theta-2\cos^2\theta)$ and is {\it expansive} for
$0\deg <\theta <\arctan\sqrt{2}\simeq 54\deg$ and $126\deg\lta\theta
<180\deg$.  On the complementary spherical sector the field is {\it
compressive}. Due to the strong non--linear dependence of the cooling
function on the gas density, it is not easy to determine without the
aid of numerical simulations which effect prevails in determining the
successive evolution of the flow, i.e., if the cooling catastrophe is
anticipated or retarded by the effect of the tidal field.  Our results
show that actually compression wins the game and cooling increases.

Successively, strong and fast oscillations in $\Lx$ occur, whose
amplitude may reach an order of magnitude. These oscillations are due
to correspondent oscillations of the ISM density in the galactic
central regions: in fact, most of the X--ray luminosity is emitted
there\footnote{Being $\Lx=\int n^2\Lambda (T)dV$, the oscillations in
$\Lx$ are basically due to variations and/or rearrangements in the hot
gas density, because $\Lambda (T)$ is nearly constant in the range of
X--ray temperatures.}, for $\Lx \propto n^2$. It is important to note
that, while these density oscillations reflect dramatically on $\Lx$,
they have little effect on the overall content of the hot gas which
decreases monotonically in this phase, as shown in Fig. 3d (medium
solid line).

\placefigure{fig4}

This strongly unsteady evolution is due to the non--spherical gas
accretion onto the galaxy central region. Converging to the center,
non--radial gas streams interact and compress each other, giving rise
to tongues of dense cold gas (see Fig. 4, where enlargements of the
gas distribution and of the velocity field in the central region are
shown). These tongues are rapidly accreted onto the galaxy
center. When the compressed gas quickly cools, $\Lx$ suddenly
increases. The cooled gas disappears in the center, leaving behind
regions of low gas density which do not radiate efficiently and tend
to expand as they are heated by SNIa: as a consequence $\Lx$ abruptly
decreases.  New streams of hot gas coming from the outer regions
contrast the expansion of the gas close to the center and form new
cold filaments.  This cycle is fed by gravity and amplified by $\Lsn$.
Actually, the hot gas mass decreases after every spike of $\Lx$
(Fig. 3d, medium solid line), while the cold gas mass increases
specularly (Fig. 3d, light solid line).  When the mass of the hot ISM
is sufficiently reduced, $\Lx$ falls by nearly two orders of
magnitude. Successively, the low density, non radiating hot gas is
able to revert the flow on large scale, and a new wind phase starts
(see Fig.3a and Fig.3d for $t\gta 12$ Gyr).

The importance of SNIa in maintaining this cycle is apparent when
considering the X--ray luminosity evolution of model MC, shown in
Fig. 5.  In this model, where $\Lsn=0$, the oscillations of $\Lx$ have
smaller amplitudes and the dramatic drop shown by the model M1 is
absent.  In fact, the $\Lx$ oscillations in model MC never exceed a
factor of three relatively to the unperturbed cooling flow solution.
Concerning the total mass budget (Fig.3d, heavy solid line), it is
apparent its substantial reduction when compared to model M0 (Fig.3d,
heavy dashed line). Thus, the tidal encounter is able to strip a
significant fraction of ISM from the M1 model.

\placefigure{fig5}

Figure 3bc show that models M2 and M3 (where the maximum approach
happens in the outflow and inflow phase, respectively) exhibit $\Lx$
oscillations similar to model M1. It is however apparent that in these
latter models the dramatic drop in $\Lx$ (occurring in model M1 at
$t\simeq 12$ Gyr), is absent over an Hubble time. This is due to the
fact that $\Lsn$ at the time of the encounter is lower than in model
M1 (Fig.3abc, dotted line). Thus less energy is available to sustain a
turbulent flow, and the $\Lx$ (and mass budget) evolution of models M2
and M3 is more similar to model MC.

The hot and cold ISM masses of models M2 and M3 (medium and light
solid lines in Fig. 3ef) evolve in the same qualitative way as in
model M1 (Fig. 3d). On the contrary, their total mass (heavy solid
line) increases, at odd with model M1. This result at first could seem
surprising because model M1 is in a wind phase for $t\gta 12$ Gyr, and
one could expect an easier degassing than in models M2 and M3.
Actually, the hot gas density in model M1 is {\it lower} than in
models M2 and M3 (as can be seen considering the amount of hot gas
mass in the three models): as a consequence, the mass loss at the
outskirts ($r=\rt$) of the galaxies {\it increases} from model M1 to
model M3.  In model M1 the mass loss rate is lower than the mass
return rate from the evolving stellar population, $\alpha (t)\mas$, the
opposite is true for models M2, and this explains the different time
behavior of the total mass in the three models.

\subsubsection{Surface Brightness}
The surface brightness profiles of all computed models do not result
strongly influenced by the tidal interaction. As a representative case
we describe the X--ray surface brightness distribution $\SX$ of model
M1 only. During all the model evolution, the isophotes remain rather
circular, although locally distorted, even for a viewing angle
$\theta=90\deg$ (i.e. a direction perpendicular to the line joining
the galaxy and the perturber, where the effect is maximum).  This can
be seen in Fig. 6, where two sets of isophotes are superimposed,
referring to $t=9.9$ Gyr (solid lines) and $t=11.1$ Gyr (dotted
lines). These times correspond to a local maximum and minimum of
$\Lx$, respectively (black dots in Fig. 3a). Radial cross
sections of $\SX$ are shown (after an angular mean) in Fig. 7.  Note
that when $\Lx$ is higher ($t=9.9$ Gyr) the extra--luminosity is
radiated mostly from the center, and so density oscillations here are
by far more important than in the rest of the galaxy.

\placefigure{fig6}

\placefigure{fig7}

\section {Discussions and Conclusions}
In this paper we have investigated the consequences of a gravitational
encounter on the dynamics of the hot, X--ray emitting ISM of cluster
elliptical galaxies.  We adopted a spherical model galaxy identical to
the King Reference Model described in CDPR in order to better compare
our results with the ``standard'' WOI picture. It turns out that once
the gas is perturbed by the gravity of a galaxy passing nearby, it
remains turbulent for the rest of the time. In fact, when the
spherical symmetry is broken by the tidal interaction, non--radial
streams converging toward the center merge thus compressing the gas,
which quickly cools and ``disappears'' into the center. Some regions
close to the center are then formed where the gas is rarefied and
tends to expand as heated by the sourcess terms. This expansion is
contrasted by the outer ISM which is falling to smaller radii, and new
cold filaments are formed.  From an observational point of view the
most striking consequences are the large oscillations of $\Lx$ coupled
to the oscillations of the gas density. These oscillations may have
amplitudes ranging over two orders of magnitude while the X--ray
surface brightness does not stray dramatically from spherical
symmetry.  Thus, at least for WOI models, the X--ray luminosity does
not represent a good diagnostic of the dynamical state of the gas when
tidal interactions are present.

The mechanism described above has been proven to be self--sustaining
and independent of the external cause which initially originated it.
In fact, we ran the model MP where, as anticipated in \S 3, the
perturber is absent, the galaxy is at rest in a inertial reference
system, and a random large--scale ``noise'' is superimposed on the
initial velocity field.  In model MP the maximum allowed relative
amplitude of the perturbation with respect to the unperturbed velocity
field is 20 per cent.  Apart from this, the initial conditions are the
same as for model M3. The mass budget and $\Lx$ evolution of model MP
are shown in Fig. 8, which should be compared with Fig. 3cf. As can be
seen, this model behaves mostly as model M3. The spikes in $\Lx$ are
somewhat larger, but this is due to our (arbitrary) choice of the
maximum amplitude of the velocity perturbations.  We point out that
model MP is rather similar to models discussed by Kritsuk et
al. (1998).  However, a significant comparison between these models
and model MP cannot be made because the latter authors stop their
simulations after only 50 Myr, a time interval too short to predict
the behaviour of the flow on cosmological times.  The major insight
derived from model MP is that {\it the evolution of perturbed WOI
models is weakly dependent on the specific nature of the
perturbation.}

We also ran a pure cooling flow model with a tidal perturbation (model
MC): in this case the $\Lx$ oscillations are significantly reduced
(approximately of a factor of ten) with respect to those exhibited by
model M3. The most important consequence is that {\it tidal
interactions cannot be at the origin of the spread in the $\Lx-\Lb$
diagram if the pure cooling flow scenario applies}.

\placefigure{fig8}

As discussed in \S2, in our simulations we neglected the cluster tidal
field, which is present in equation (5).  Although the magnitude of the
cluster tidal field during the encounter is negligible (see \S 2.1.1),
one can ask whether the {\it cumulative} effects of such field over
cosmological times are significative.  In order to check this
possibility, as anticipated in \S 3 we ran the model MT, in which
$\Lsn=0$, no perturber is present, and only the cluster tidal field is
considered.  The resulting evolution of this model (not shown here) is
very similar to that of model MC, but the amplitudes in the $\Lx$
oscillations are much reduced. We thus conclude that the cluster tidal
field can be safely neglected when studying the effects of
galaxy--galaxy encounters like those considered here on the X--ray
properties of early--type galaxies.

Moreover, it must be added that the role of the cluster tidal field is
further reduced by the fact that during a Hubble time a galaxy may
experience several encounters of the type discussed here. A simple
estimate of this number can be obtained as follows.  Suppose that the
galaxy belongs to a typical rich cluster like that described in \S
2.1.1, and containing $N_{\rm T}=300$ galaxies, whose luminosity
function is given by
\begin{equation}
\phi(L)dL=N_{\star}\left({L\over\Lstar}\right)^{-\alpha}
          \exp{(-L/\Lstar)}d\left({L\over\Lstar}\right),
\end{equation}
where $\Lstar\simeq\,4.74\times\,10^{10}$ $\lsun$, $\alpha=5/4$
(Schechter 1976, Sarazin 1986), and with lower and upper limits of
$10^8\lsun$ and $\infty$, respectively.  Using simple geometrical
arguments it is easy to show that, for an impact parameter less than
or equal to $\rpmin$, the {\it mean} number of interactions per galaxy
during an Hubble time $\tH$ is given by
\begin{equation}
n_{\rm i}\simeq {\sigcl\tH\over\rcl}
                \left({\rpmin\over\rcl}\right)^2
                {\kappa N_{\rm T}\over 10}.
\end{equation}
In the equation above $\kappa$ is the fraction of the total number of
galaxies with luminosity greater than $\Lb$. For example,
$\kappa\simeq 0.01$ for $\Lb=5\times 10^{10}\lsun$ (the luminosity of
our galaxy model); $\kappa\simeq 0.1$ for $\Lb= 1.6\times
10^{10}\lsun$ (one third of the luminosity of our galaxy model).  It
turns out that $n_{\rm i}\simeq 2$ for $\kappa=0.01$.  The same result
is obtained considering galaxies with $\Lb\geq 1.6\times 10^{10}\lsun$
in a poor cluster (or loose groups), with $N_{\rm T}=30$, $\rcl =200$
kpc and $\sigcl= 250$ km s$^{-1}$ (see, e.g., Bahcall 1998).

Obviously the above analysis does not apply to small compact galaxy
groups containing only a few members.  In fact, these systems are
highly collisional, and the mean separation between galaxies is
comparable to their dimensions (see, e.g., Kelm, Focardi, \& Palumbo
1998). As a consequence, strong gravitational interactions are
frequent, the evolution of gas flows is higly perturbed,
and the results of our simulations can not be directly applied in this
case. 

From a more observational point of view, the importance of the
discussion above is also strengthen by the fact that, while it is true
that the majority of elliptical galaxies are found in clusters, the
galaxies from which the $\Lx -\Lb$ diagrams are constructed are either
in groups or in the Virgo cluster (which is less rich than the cluster
assumed in our simulations). Hovewer, from the discussion above it
results that the ``turbulent'' mechanism here described should be
quite common in ellipticals in clusters and groups, while galaxies in
compact groups can hardly avoid strong gravitational interactions,
that should produce either strong variations in their X--ray
luminosity or even substantial degassing.

As pointed out in \S 2.2, with the computed models available, we are
now in the position to evaluate the error induced by neglecting the
non--inertial forces associated with the rotation of the adopted
reference frame.  From an analysis of the model's velocity field at
different times, it turns out that, when the perturber is closest, the
Coriolis and centrifugal forces are larger than the gravitational
forces only within a small volume (with a characteristic size of $\sim
2$ kpc) at the edge of the galaxy facing the perturber, where the
total gravity is very low. Later ($t-\tm\gta 200$ Myr), the gravity
largely prevails (by a factor 100--1000) all over the computational
grid.  This is mainly due to the fact that for the assumed orbit
$||\Ome||\propto (t-\tm)^{-2}$ (see equation [10]), and so it quickly
reduces after the encounter; moreover, the non inertial force
associated with $\dot\Ome$ vanishes when $\rp=\rpmin$, and, for
$|t-\tm|> \rpmin/V\simeq 100$ Myr, $||\dot\Ome||\propto |t-\tm|^{-3}$.
As a consequence, we believe that our results are reliable, and,
although neglecting the fictitious forces due to rotation may prevent
from obtaining the exact values of the $\Lx$ spikes, the essential
physics is captured by the computed models.

In conclusion, the main results of this work can be summarized as follows:
\begin{itemize}

\item In WOI models the $\Lx$ oscillations induced by tidal encounters last
for several Gyrs after the encounter is over. The qualitative
evolution of these oscillations is not strongly dependent on the flow
phase (wind, outflow, or inflow) occurring when the encounter starts;
moreover, they appear to be rather independent of the physical process
generating them. The characteristic period of such oscillations is of
the order of the sound crossing time of the hot gas in the galactic
inner region. If the encounter happens quite early, when $\Lsn$ is
still relatively high, then a catastrophic drop of $\Lx$ is produced.
These oscillations can contribute to the observed spread in the $\Lx
-\Lb$ diagram for galaxies in clusters and groups.  Of course, this
mechanism cannot be at the origin of the $\Lx$ scatter in isolated
galaxies.

\item When compared to WOI models, the amplitude of the $\Lx$ oscillations
in cooling flow models (with $\Lsn=0$) is substantially reduced. As a
consequence, tidal interactions cannot be at the origin of the
observed spread in the $\Lx -\Lb$ diagram in the ``pure'' (no SNIa
heating) cooling flow scenario.

\item In all computed models, the X--ray surface brightness $\SX$ does not 
appear to be strongly distorted by tidal interactions. However, the 
(angular mean) radial profile of $\SX$ is steeper in the inner region 
in coincidence with the $\Lx$ spikes. 

\item We found that the tidal stripping due to the perturber is effective.
In fact, approximately half of the gas content of the galaxy is lost
in perturbed WOI models when compared to the unperturbed one. In our
pure cooling flow model the fraction of lost mass is reduced to one
fourth.

\item The cluster tidal field {\it alone} appears to be ineffective 
in producing large $\Lx$ oscillations when compared to tidal
interactions. Moreover, during a Hubble time, galaxies in clusters
suffers several encounters of the kind described here, and so the
importance of the cluster tidal field on the X--ray properties of
early--type galaxies is further reduced.

\end{itemize}

\acknowledgments
We would like to thank Neta Bahcall, Giuseppe Bertin, James Binney,
and Silvia Pellegrini for very useful discussions, and Giovanna Stirpe
and Steven Shore for a careful reading of the manuscript.  We also
acknowledge the anonymous referee for very constructive comments.
This work was partially supported by Italian MURST, contract CoFin98,
and by ASI contracts, ASI-95-RS-152 and ASI-ARS-96-70.

\appendix

\section{The Hydrodynamical Equations with General Source Terms in
         Non--Inertial Reference Systems}
In Appendix A.1 we present the treatment of mass, momentum, and energy
source terms in hydrodynamics, in the general anisotropic case.  In
Appendix A.2 the hydrodynamical equations with general source terms
are expressed in a non inertial reference frame.
\subsection{The Source Terms}
In the following treatment we assume the presence in the (inertial)
reference system $S$ of a single source field, the generalization to
the case of more than one field being straightforward.  $\xv=\xci\evi$
and $\vv=\vci\evi$ are the position and the velocity vectors in $S$,
respectively.  Let the {\it source field} be described by a 
distribution function $f=f(\xv,\vv;t)$ (see, e.g., BT), so that
$n(\xv;t)=\int_{\Re^3}f\,\dtv$ represents the number density of
sources.

The {\it mass return} associated with the source field is given by 
the function $m=m(\xv,\vv,\nv;t)$, 
where $\nv=\nci\evi$ is a unitary vector accounting for the
possibility of {\it anisotropic} mass sources; the physical units of
$m$ are mass for unit time for unit solid angle.  The total mass return 
per unit time, volume and solid angle associated with the source field is
then given by $\mu(\xv,\nv;t)=\int_{\Re^3}m\,f\,\dtv$.
Finally, integrating over the whole solid angle, we obtain 
the total mass return rate per unit time and volume at $\xv$:  
\begin{equation}
\sourceM(\xv;t)=\int_{4\pi}\mu\,\ddn.
\end{equation}
For example, for mass sources independent of the source velocity
$\vv$, $\mu(\xv,\nv;t)=nm$; moreover for an isotropic source field,
$\sourceM(\xv;t)=4\pi nm$.

The {\it momentum return} associated with the source field is
given by the vectorial function
$\pv=m(\xv,\vv,\nv;t)[\vv+\us(\xv,\vv,\nv;t)\nv]$,
where $\us$ is the modulus of the velocity of the material associated
with the source field along $\nv$ {\it with respect to the source
velocity $\vv$}.  The total momentum per unit time, volume and solid
angle associated with the source field is then given by
$\pgv(\xv,\nv;t)=\int_{\Re^3}\pv\,f\,\dtv$.
Finally, integrating over the whole solid angle, 
\begin{equation}
\sourcePv(\xv;t)=\int_{4\pi}\pgv\,\ddn.
\end{equation}
In particular, for mass sources and ejection velocity independent of
the source velocity $\vv$, $\pgv(\xv,\nv;t)=nm [\vm+\us\nv]$, where
\begin{equation}
n(\xv;t)\vm (\xv;t)=\int_{\Re^3}\vv\,f\,\dtv
\end{equation}
is the {\it source streaming velocity field}, $\vm=\vmci\evi$ (see,
e.g., BT).  Moreover, if $m$ and $\us$ are also isotropic,
$\sourcePv(\xv;t)=\sourceM\vm$.

The {\it energy source} associated with the source field is made of the
contributions of two distinct parts, i.e., the {\it internal energy
source} and the {\it kinetic energy source}. The internal energy
source is described (in analogy with the mass source) by the function
$e=e(\xv,\vv,\nv;t)$.  The physical units of $e$ are energy per unit
time per unit solid angle per unit mass.  The total internal energy
per unit time, volume and solid angle associated with the source field
is then given by $\eps(\xv,\nv;t)=\int_{\Re^3}m\,e\,f\,\dtv$.
Finally, integrating over the whole solid angle,
\begin{equation}
\sourceE(\xv;t)=\int_{4\pi}\eps\,\ddn.
\end{equation}
In particular, for mass and internal energy sources independent of the
velocity source $\vv$, $\eps(\xv,\nv;t)=\mu e$.  Moreover, in case of
isotropy, $\sourceE(\xv;t)=\sourceM e$.  The kinetic energy source is
given by $k={1\over
2}m(\xv,\vv,\nv;t)||\vv+\us(\xv,\vv,\nv;t)\nv||^2$.  The total kinetic
energy per unit time, volume and solid angle associated with the source
field is then given by $\kappa(\xv,\nv;t)=\int_{\Re^3}k\,f\,\dtv$.
Finally, integrating over the whole solid angle,
\begin{equation}
\sourceK(\xv;t)=\int_{4\pi}\kappa\,\ddn.
\end{equation}
In particular, for mass sources and ejection velocity independent of
the velocity source $\vv$, $\kappa(\xv,\nv;t)=nm [||\vm||^2+\Tr
(\sigma^2)+\us^2+ 2\us<\nv,\vm>]/2$, where $\Tr (\sigma^2)$ is the
trace of the velocity dispersion tensor associated with the source
motion:
\begin{equation}
n(\xv;t)\sigij(\xv;t)=\int_{\Re^3}(\vci-\vmci)\,(\vcj-\vmcj)\,f\,\dtv.
\end{equation}
Moreover, if $m$ and $\us$ are also isotropic,
$\sourceK(\xv;t)=\sourceM [||\vm||^2+\us^2+
\Tr (\sigma^2)]/2$.  {\it Note that a source field is called isotropic
only if the functions $m$, $\pv$, $\us$ and $e$ are all independent of
$\nv$.}

\subsection{The Hydrodynamical Equations in a Non--Inertial Reference System}
The basic hydrodynamical equations with general source terms are here
derived (in the inviscid case) by using the standard approach based on
the {\it Reynolds Transport Theorem}. In fact, this formulation is
particularly useful when moving to a non--inertial reference frame
(see, e.g., Narasimhan 1993).

As usual, we denote with $D/Dt=\partial/\partial
t+\vci\partial/\partial\xci$ the Lagrangian derivative when expressed
in Eulerian form. After some manipulation of the conservation laws
expressed in integral form, and application of the Transport Theorem,
one obtains the {\it continuity} equation
\begin{equation}
{D\rho\over Dt}+\rho\gradx\cdot\uv=\sourceM.
\end{equation}
The three components of the {\it momentum} equation are given by
\begin{equation}
\rho{D\uv\over Dt}=\rho\gv-\gradx p +\sourcePv-\sourceM\uv,
\end{equation}
where $\gv$ is the total physical acceleration and $p=p(\xv;t)$ is the
{\it thermodynamical pressure}. 
Finally, the {\it energy} equation is derived:
\begin{equation}
{DE\over Dt}+(E+p)\gradx\cdot\uv=\sourceE+\sourceK+
                                 {\sourceM\over 2}||\uv||^2-
                                 <\sourcePv,\uv>-\sourceL.
\end{equation}
where $E$ is the {\it internal energy} per unit volume, and $\sourceL$
describes the radiative losses per unit volume and unit time. Note
that thermal conduction is not considered here.

In case of a completely isotropic source field (the case of our model
galaxies), the momentum and energy equations can be rewritten as:
\begin{equation}
\rho{D\uv\over Dt}=\rho\gv-\gradx p+\sourceM(\vm-\uv).
\end{equation}
and
\begin{equation}
{DE\over Dt}+(E+p)\gradx\cdot\uv=
              {\sourceM\over 2}||\uv-\vm||^2+
              \sourceM\left[e+{\us^2\over 2}+
                              {\Tr (\sigma^2)\over 2}\right]-
              \sourceL.
\end{equation}

We move now from $S$ to a non--inertial frame $S'$. As well known, the
position, velocity, and acceleration vectors $\xv$, $\dot\xv$,
$\ddot\xv$ (in $S$)
are related to the corresponding vectors in $S'$ by
\begin{equation}
\xv=\Rv+\Om\csiv,
\end{equation}
\begin{equation}
\dot\xv=\Vv+\Om (\dot\csiv+\Ome\wedge\csiv),
\end{equation}
\begin{equation}
\ddot\xv=\Av+\Om [\ddot\csiv+2\Ome\wedge\dot\csiv+\dOme\wedge\csiv+
                 \Ome\wedge(\Ome\wedge\csiv)],
\end{equation}
where $\Rv(t)$, $\Vv(t)$ and $\Av(t)$ are the position, velocity, and
acceleration of the origin of $S'$ (see, e.g., Arnol'd 1980).  $\Om(t)
\in SO(3)$ is a rotation matrix, and $\Ome$ is the dual of the
antisymmetric matrix $\Omt\dot\Om$.  Geometrically, $\Ome$ is the
angular velocity of $S'$ with respect to $S$ resolved along the basis
of $S'$.  When resolved in $S$, we have $\ome=\Om\Ome$.

Now, the scalar continuity and energy equations are transformed in the
corresponding equations in $S'$.  First, the relation between the
velocity field $\uv(\xv;t)$ (in $S$) and $\nuv(\csiv;t)$ (in $S'$) is
obtained from the definition of velocity field as Lagrangian
derivative of the position vector, and so from equation (A13)
$\uv(\xv;t)=\Vv+\Om [\nuv(\csiv;t)+\Ome\wedge\csiv]$. Second, in $S'$,
the expression for the Lagrangian derivative is
$D/Dt=\partial/\partial t+\nuci\partial/\partial\csici$, moreover, it
is easily proved that $\gradx\cdot\uv =\gradcsi\cdot\nuv$.

The transformation of continuity and energy equations is now
immediate; some care is required for the momentum equation, when
considering the Lagrangian derivative of the velocity.  The final
equations (for isotropic sources) in conservative form are
\begin{equation}
{\partial\rho\over\partial t}+\gradcsi\cdot\rho\nuv=\sourceM,
\end{equation}
\begin{equation}
{\partial\rho\nuv\over\partial t}+\gradcsi\cdot (\rho\nuv\otimes\nuv)=
                    \rho\Omt(\gv-\Av)-
                    \rho[2\Ome\wedge\nuv+\dOme\wedge\csiv+
                         \Ome\wedge(\Ome\wedge\csiv)]-
                    \gradcsi p+
                    \sourceM\vm ',
\end{equation}
\begin{equation}
{\partial E\over\partial t}+\gradcsi\cdot E\nuv=
                     -p\gradcsi\cdot\nuv+
                     {\sourceM\over 2}||\nuv-\vm '||^2+
                     \sourceM\left[e+{\us^2\over 2}+
                     {\Tr (\sigma^2)\over 2}\right]-
                     \sourceL .
\end{equation}
\clearpage

\clearpage

\figcaption[]{ The modulus of the dimensionless radial (solid line) and 
               tangential (dashed line) component of the cluster tidal field,
               as a function of the cluster radius. The normalization 
               constant is $|\Tp|^{\rm min}=G\map/\rp^3$, for $\rp=100$ kpc.
               The explicit expression of the $q$ function entering equation 
               (8) can be found in Ciotti \& Dutta (1994).
               The cluster mass is obtained by using the standard relation 
               $4\pi G\rocl (0)\rcl^2=9\sigcl^2$.
              \label{fig1}}

\figcaption[]{ Logarithmic density distribution and velocity field of the gas
               in model M1 at different times. The perturber is on the right.
              \label{fig2}}

\figcaption[]{ The upper panels show the evolution of $\Lx$ for the models
               M1, M2, and M3 (solid lines), for the model M0 (dashed lines), 
               and $\Lsn$ (dotted lines). The two dots refer to the times 
               indicated in Fig. 2bc, and Figs. 6 and 7.
               The evolution of the total gas mass (heavy lines), 
               hot gas mass (medium lines), and cold gas mass (light lines) 
               are shown in the lower panels. Dashed lines refer to model M0, 
               solid lines to models M1, M2, and M3.
              \label{fig3}}

\figcaption[]{ Logarithmic density distribution and velocity field of the gas
               in model M1 near the galactic center at two different times 
               (in particular the left panel is the enlargment of the central
               region of Fig. 2b). 
               The turbulent regime of the flow, and the transient 
               cold filaments are apparent.
              \label{fig4}}

\figcaption[]{ Time evolution of 
               $\Lx$ for the cooling flow, tidally perturbed MC model 
               (solid line). $\Lx$ of the same model without any perturbation
               (dashed line) is shown for comparison. 
              \label{fig5}}

\figcaption[]{ X--ray surface brightness $\SX$ of model M1 at $t=9.9$ 
               Gyr (solid curves) and $t=11.1$ Gyr (dotted curves). 
               These times are the same as in panels $b$ and $c$ of Fig. 2.
               The perturber is located on the right. Labels are in kpc.
              \label{fig6}}

\figcaption[]{ Radial distribution of the angular averaged surface brightness 
               $\SX$ shown in Fig. 6. Solid and dotted lines correspond to 
               a $\Lx$ maximum and minimum, respectively (see the black dots 
               in Fig.3a). 
               $r_e =\sqrt{\rt\rcl}\simeq 4.9$ kpc is the effective radius of 
               the stellar density distribution. $\SX^{\rm max}$ is the central
               value of the X--ray surface brightness of model M0.         
              \label{fig7}}

\figcaption[]{ X--ray luminosity $\Lx$ and gas mass evolution for the model 
               MP. In the upper panel the dotted line represents $\Lsn$.
               In the lower panel the evolution of the total gas mass 
               (heavy line), hot gas mass (medium line), and cold gas mass 
               (light line) is shown.
               \label{fig8}}

\clearpage
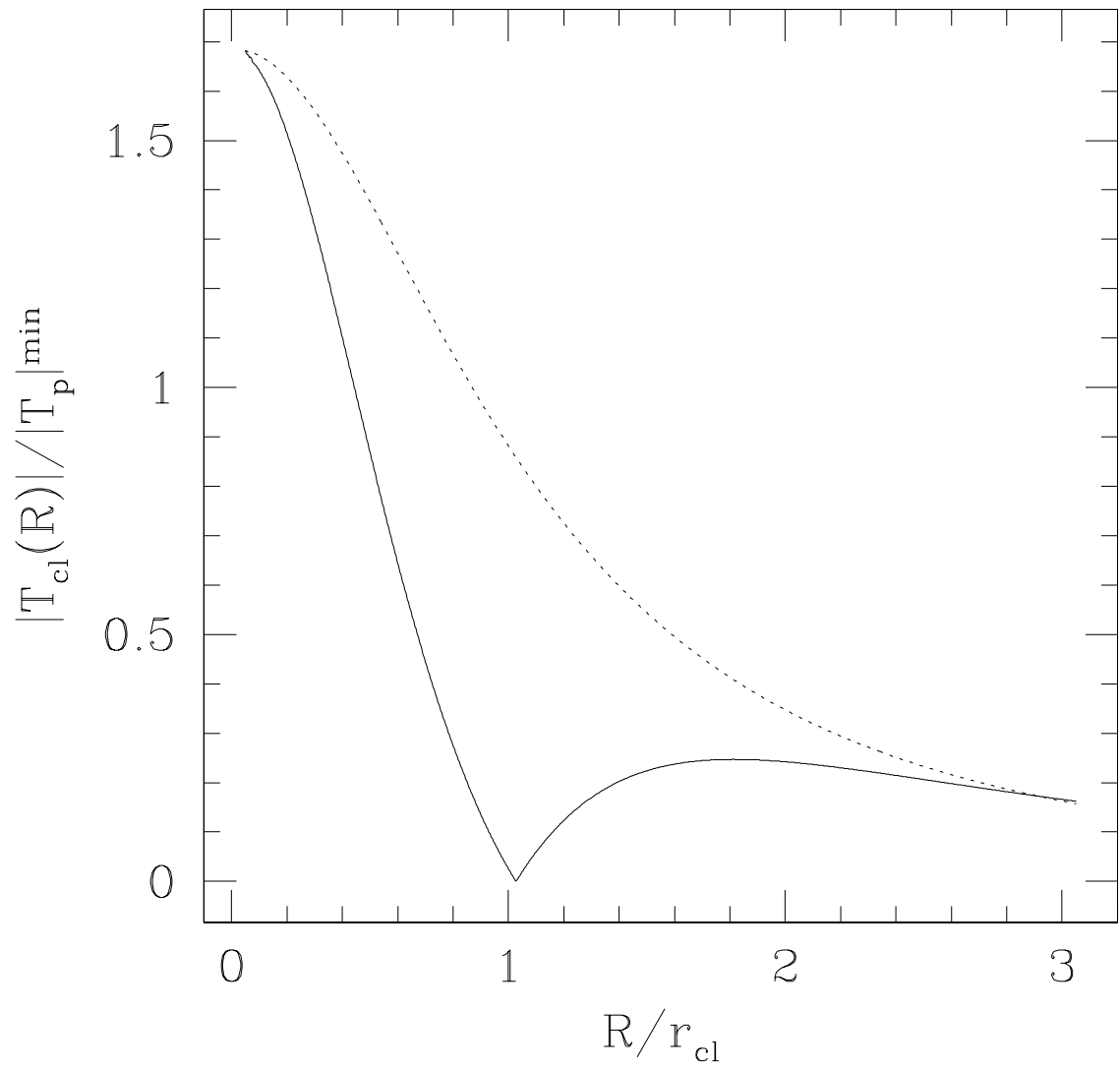
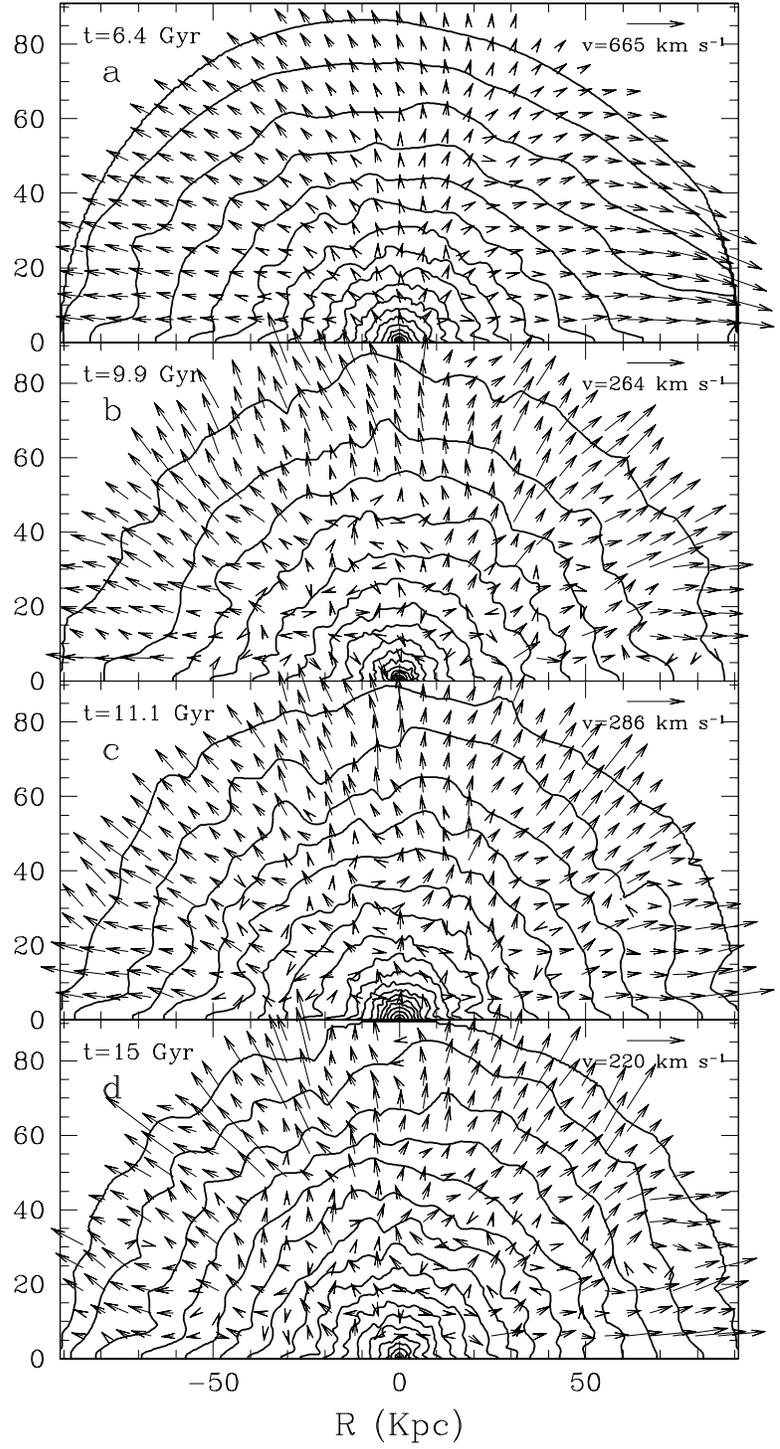
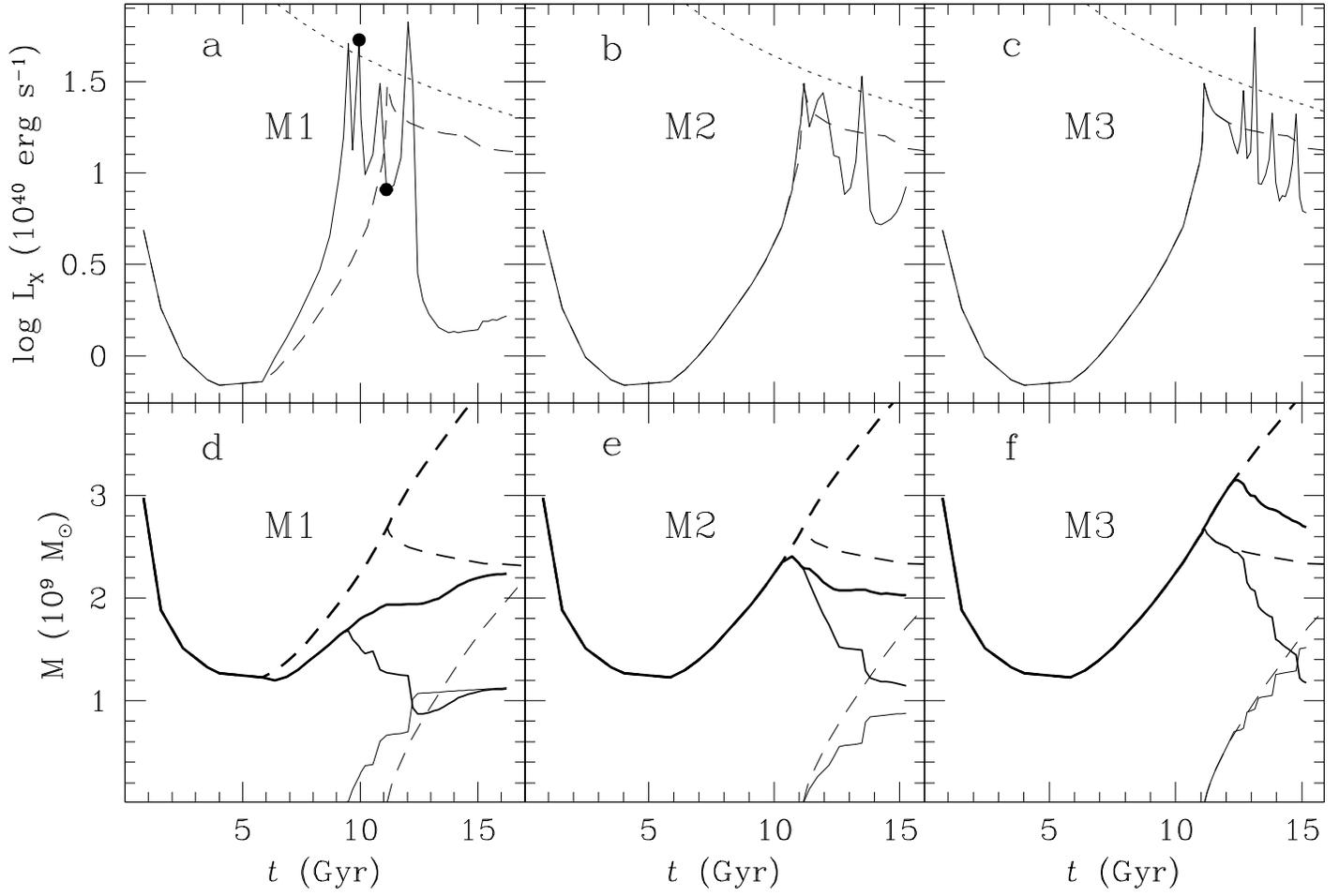
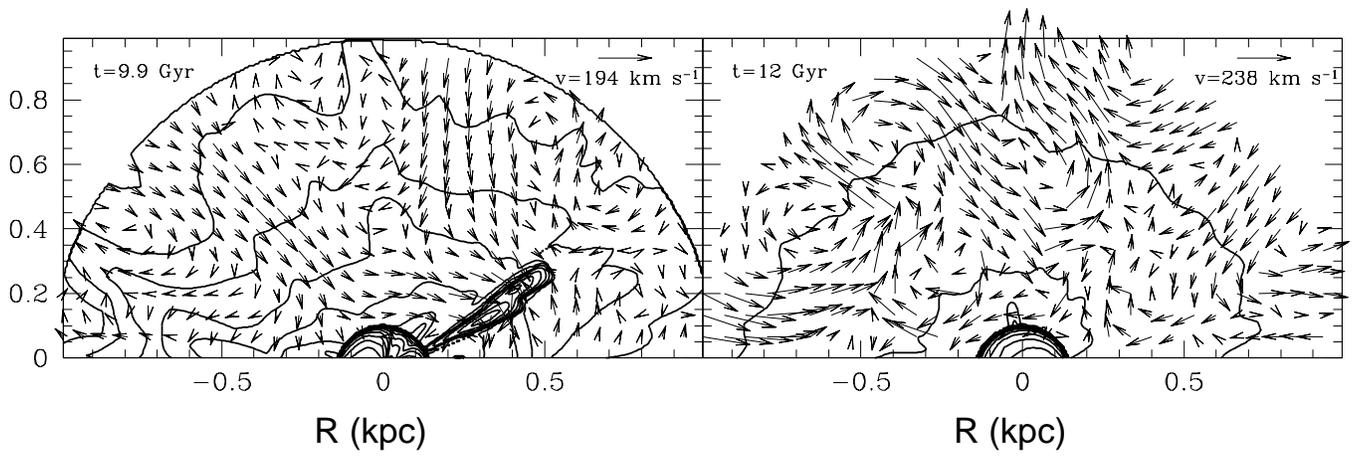
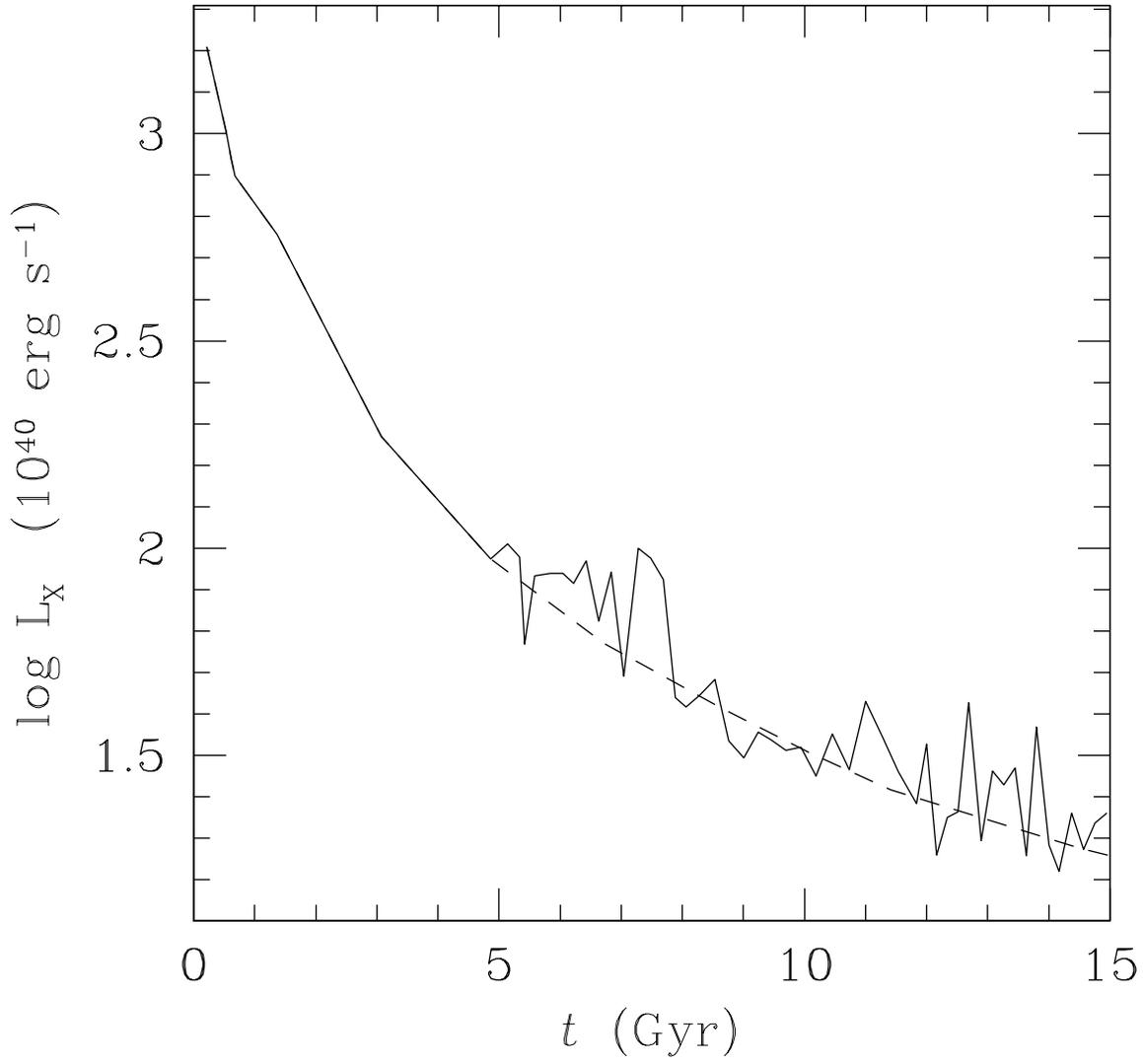
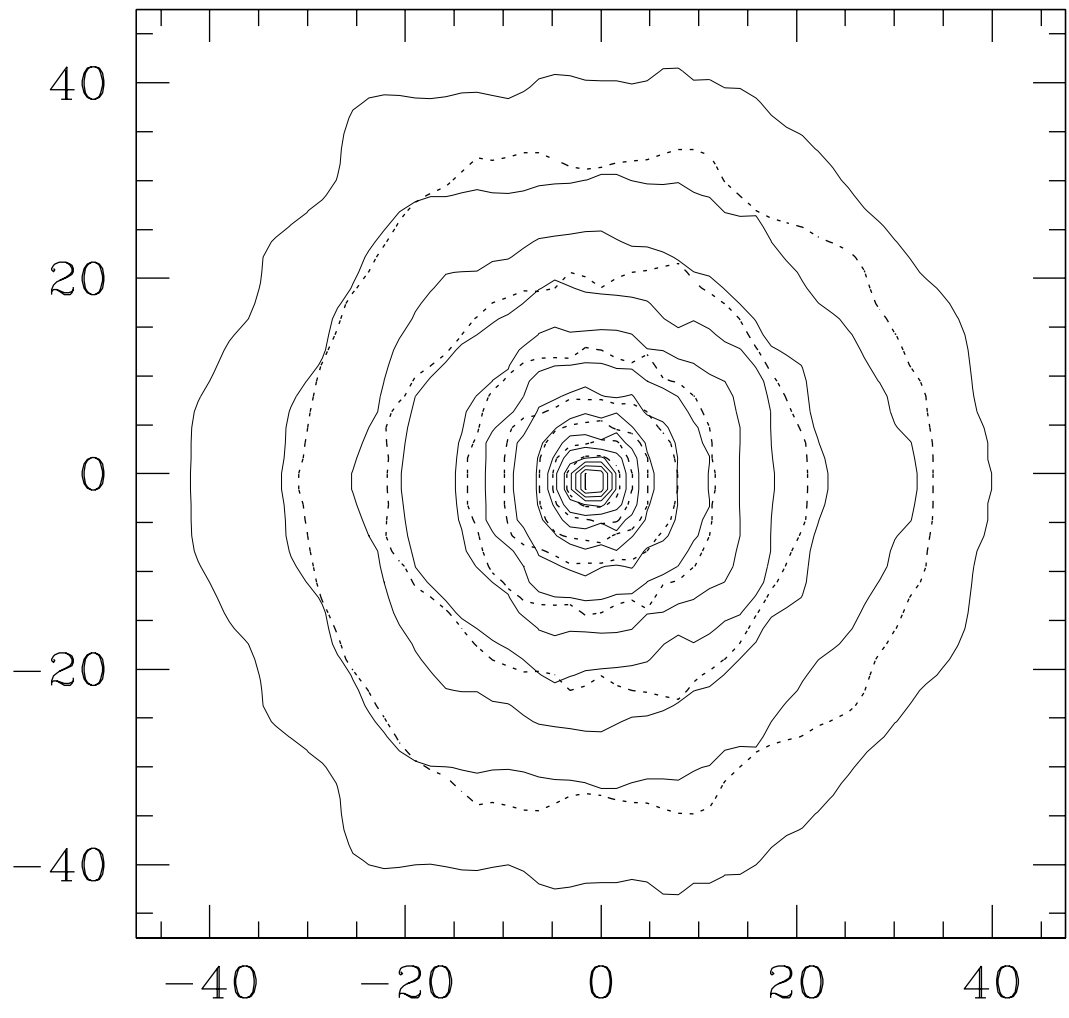
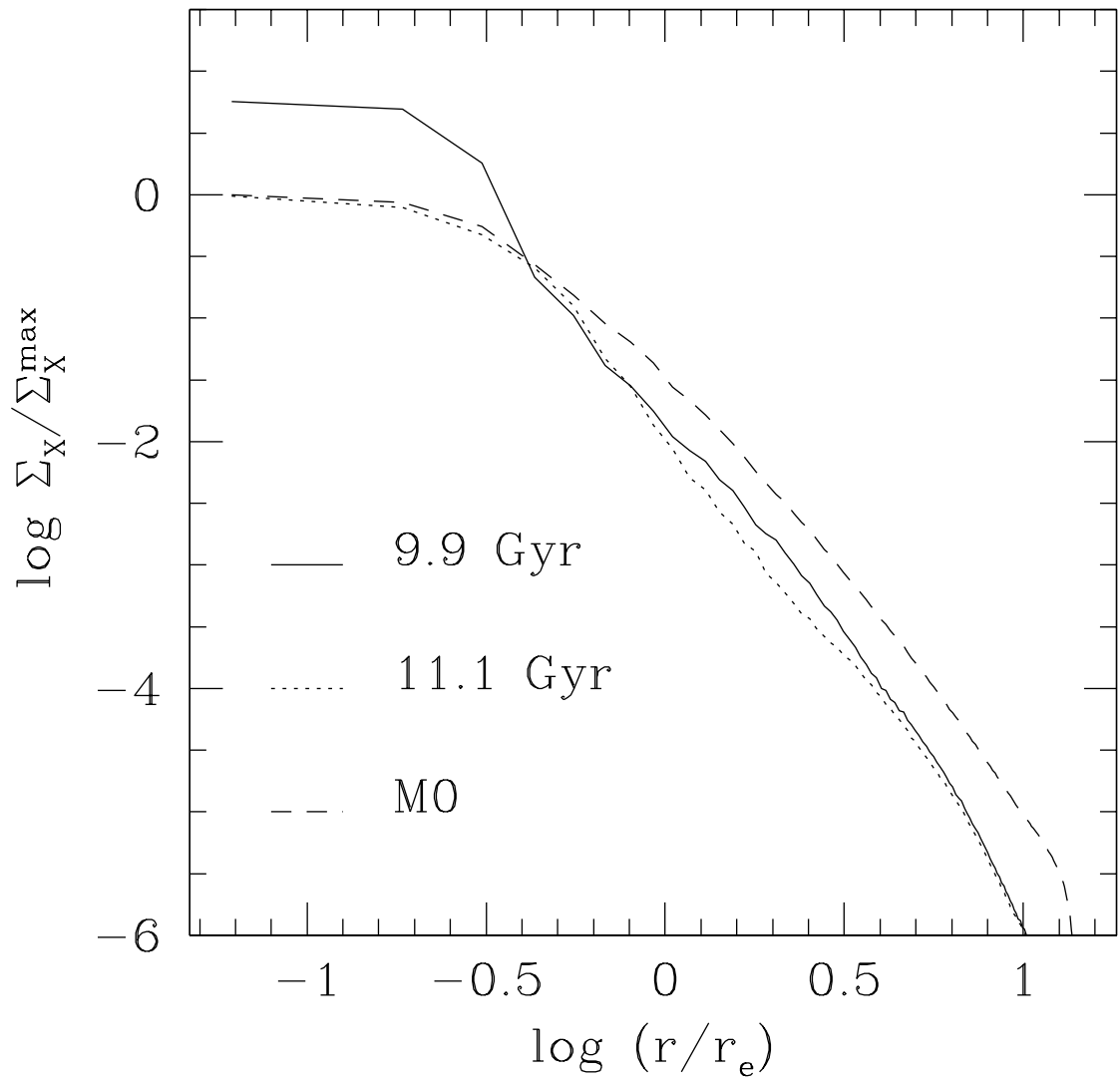
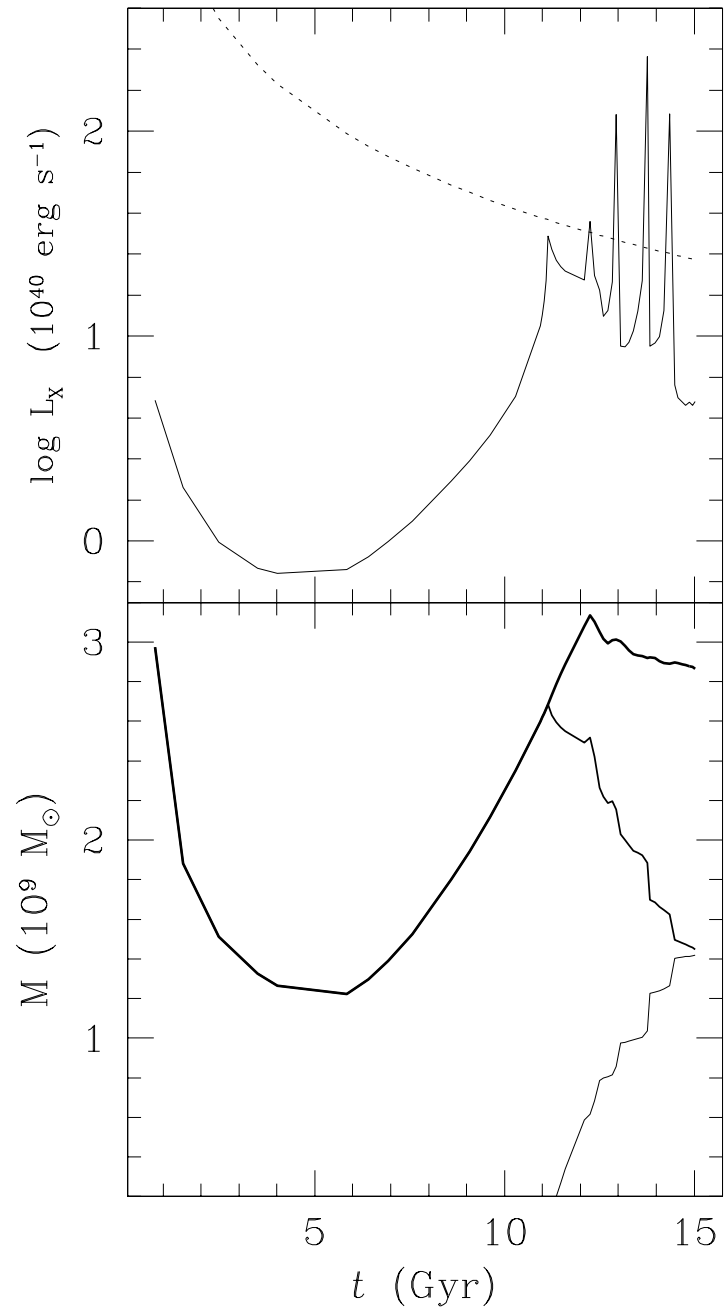
\end{document}